\documentclass[12pt]{article}
\usepackage{bbm}
\usepackage{amsfonts}
\usepackage{mathrsfs}
\usepackage {amssymb}
\usepackage {amsmath}
\usepackage{amsthm}
\usepackage{latexsym}
\usepackage{indentfirst}
\usepackage{enumitem}

\newcommand{\bbF}{\mathbb{F}}
\newcommand{\F}{\mathbb{F}}

\newcommand{\Z}{\mathbb{Z}}

\usepackage{lastpage}
\usepackage{epsfig}

\usepackage{anysize}
\marginsize{3cm}{3cm}{2cm}{2cm} \normalsize

\newtheorem{thm}{\bfseries  Theorem}[section]
\newtheorem{lem}[thm]{\bfseries   Lemma}
\newtheorem{coro}[thm]{\bfseries   Corollary}

\newtheorem{Def}[thm]{\bfseries  Definition}
\newtheorem{ex}[thm]{\bfseries   Example}
\newtheorem{rem}[thm]{\bfseries   Remark}
\newtheorem{open}[thm]{\bfseries   Open Problem}
\newtheorem{prop}[thm]{\bfseries   Proposition}
\begin{document}

\title{ Trace codes over  $\Z_4$ and Boolean function}

\date{}
\author{ Minjia Shi\footnote{Key Laboratory of Intelligent Computing $\&$ Signal Processing
Ministry of Education, Anhui University No. 3 Feixi Road, Hefei Anhui Province 230039, P. R. China, National Mobile Communications Research Laboratory, Southeast University and School of Mathematical Sciences of Anhui University,
Anhui, 230601, P. R. China.}, Yan Liu, Randriam Hugues, Lin Sok, Patrick Sol$\acute{e}$\footnote{CNRS/LAGA  Universit\'e Paris 8, 2 rue de la Libert\'e, 93 526 Saint-Denis, France.}
}

\maketitle

\begin{abstract} We construct trace codes over $\Z_4$
by using Boolean functions and skew sets, respectively. Their Lee weight distribution is studied by using a Galois ring version of the
Walsh-Hadamard transform and exponential sums. We obtain a new family of optimal two-weight codes over $\Z_4.$
\end{abstract}


\section{Introduction}
Most trace codes have their coordinate sets indexed by the elements of a finite field, or in the case of $\Z_4$-codes by the Teichm\"uller set of a Galois ring \cite{KH}.
This is the case, for instance, of the quaternary Kerdock codes \cite{HKCSS}.
 Recently, Ding
investigated a different way of constructing trace codes by indexing their coordinate places by the elements of a difference set \cite{D},
and in some cases, the support of a Boolean function \cite[\S VI]{D}.

In this paper we generalize Ding's approach to $\Z_4$-codes. The support $S_f$ of a Boolean function $f$ is mapped to a subset of a Teichm\"uller set by inverse reduction modulo $2.$
The Lee weight distribution of the code is studied by means of a variant of the Walsh-Hadamard transform, which is in fact the Walsh-Hadamard transform for a family of generalized
Boolean function with domain $\F_2^n$ and range $\Z_4.$ In fact some of our codes (\S 4.1) will have the codes of \cite{D} as residue codes and as torsion codes. Another approach, with a different defining set yields
three-weight codes. Using a natural generalization of skew sets of \cite{D}, we obtain two-weight codes the Gray images of which meet the recent Griesmer bound for nonlinear binary codes of \cite{BGMS}.

The material is organized as follows. The next section sets up the basic notations and definitions. Section 3 gives a character sum approach to the weight distribution of our trace codes.
Section 4 discusses the two families of codes we mentioned and give, for the second family, its weight distributions when the Boolean function is bent or semi-bent.
Section 5 recapitulates the obtained results and makes some conjectures for future research.\\

\section{Preliminaries}

\subsection{Rings}
In this subsection, we recall several basic facts on the algebraic structure of the Galois ring $GR(4,m)$
and fix several
basic notations. For more knowledge on Galois rings we refer to Wan's book \cite{W}.

For convenience later, let ${\cal R}=GR(4,m).$ Denoted by ${\cal R}^*$ the group of units. In ${\cal R}$ with maximal ideal $I=\langle 2\rangle$, there exists a nonzero element $ \xi$ of order $2^m-1$, which
is a root of a basic primitive polynomial $h(x)$ of degree $m$  over $\mathbb{Z}_4$ and
${\cal R}=\mathbb{Z}_4[\xi]$. Let ${\cal T}=\{0,1,\xi,\xi^2,\ldots, \xi^{2^m-2}\}$. It can be showed that any element $c\in {\cal R}$ can be written uniquely as $c=a+2b$ with $a,b\in {\cal T}$. It can be also showed that ${\cal T} \equiv \bbF_{2^m} \pmod{2}$.
%
%
%

 Recall that the {\em trace} map $tr$ from $\bbF_{2^m}$ to $\bbF_2$ is defined by
 $tr(a)=\sum\limits_{i=0}^{m-1}a^{2^i}$
for all $a\in \bbF_{2^m}$. Define the {\em generalized trace} map $Tr$ from ${\cal R}$ to $\mathbb{Z}_4$
by $Tr(c)=tr(a)+2tr(b)$ for all $c=a+2b\in {\cal R}.$

\subsection{Codes and Gray map}
 A {\em linear code} $C$ over $\Z_4$ of length $n$ is a $\Z_4$-submodule of $\Z_4^n$. For any two vertors $\mathbf{x}=(x_1,x_2,\ldots,x_n)$, $\mathbf{y}=(y_1,y_2,\ldots,y_n)\in\Z_4^n$, the Euclidean inner product of $\mathbf{x}$ and $\mathbf{y}$
  is defined by $\langle \mathbf{x},\mathbf{y}\rangle=\sum\limits_{i=1}^nx_iy_i$, where the operation is performed in $\Z_4$. The dual code of $C$ is denoted by $C^\perp$ and
  defined as $C^\perp=\{\mathbf{y}\in \Z_4^n:\langle \mathbf{x},\mathbf{y}\rangle =0, \forall \mathbf{x}\in C\}.$ By definition, $C^\perp$ is also a linear code over $\Z_4$.
  The {\em residue code} $Res(C)$ of $C$ is the binary code defined as $Res(C)=\{ \mathbf{x}\in \F_2^n :\, \exists \mathbf{y} \in C, \mathbf{y} \equiv \mathbf{x} \pmod{2}\}.$
  The {\em torsion code} $Tor(C)$ is the binary code defined as $Tor(C)=\{ \mathbf{x}\in \F_2^n :\, \exists \mathbf{y} \in C,\, \mathbf{y}=2\mathbf{x}\}.$
  The {\em Lee weight} $w_L(\mathbf{x})$ of $\mathbf{x}=(x_1,x_2,\ldots,x_n)$ is defined as $w_L(\mathbf{x})=n_1(\mathbf{x})+2n_2(\mathbf{x})+n_3(\mathbf{x})$,
  where $n_i(\mathbf{x})$ denote the number of occurences of a $i$ symbol in $\mathbf{x}$.

For any $\mathbf{x}=(x_1,x_2,\ldots,x_n)\in \Z_4^n$ with
$x_i=r_i+uq_i$, the {\em Gray map} $\phi$ from $\Z_4^n$ to $\mathbb{F}_2^2$ is given by $\phi(\mathbf{x})=(q(\mathbf{x}),r(\mathbf{x})+q(\mathbf{x})) $, where
$r(\mathbf{x})=(r_1,r_2,\ldots,r_n),q(\mathbf{x})=(q_1,q_2,\ldots,q_n)$
are binary vectors. Then $\phi$ is a weight-preserving map from ($\Z_4^n$, Lee weight) to
($\mathbb{F}_2^{2n}$, Hamming weight), that is, $w_L(\mathbf{x})=w_H(\phi(\mathbf{x}))$, where $w_H(\phi(\mathbf{x}))$ denotes
the number of nonzero positions in the binary vector $\phi(\mathbf{x}).$\\

%
\section{Trace codes}

\subsection{ Description of Trace Codes}
Let $D=\{d_1,d_2,\ldots,d_n\}\subseteq {\cal R}\backslash \{0\}$. We define a linear code of length $n$
over $\mathbb{Z}_4$ by $C_D=\{c_a=(Tr(ad_1),Tr(ad_2),\ldots,Tr(ad_n) ):a\in {\cal R}\},$
and call $D$ the defining set of this code $C_D$. This construction is generic in the sense that many classes
of known codes could be produced by selecting the defining
set \cite{D}. The objective of this paper is to construct linear codes $C_D$
by using three classes of $D'$s defined later. If the set $D$ is well
chosen, the code $C_D$ may have good or optimal parameters.
Otherwise, the code $C_D$ could have bad parameters.

\subsection{ The Weights of $C_D$}

It is convenient to define for each $b\in {\cal R}$, $c_b=(Tr(bd_1),\ldots,Tr(bd_n))$.
Let $N_b(j)=|\{1\leq i\leq n: Tr(bd_i)=j\}|$  for $0\leq j\leq 3$. Then the Lee weight $w_L(c_b)$ of $c_b$ is $n-N_b(0)+N_b(2)$ for
each $b\in {\cal R}$.

It is easily seen that for any $D=\{d_1,d_2,\ldots,d_n\}\subseteq {\cal R}\backslash \{0\}$,
we have \begin{eqnarray*}
         N_b(j) &=& \frac{1}{4}\sum_{i=1}^n\sum_{y=0}^3i^{y[Tr(bd_i)-j]} \\
           &=&\frac{1}{4}\Big[n+ \sum_{i=1}^n\sum_{y=1}^3i^{y[Tr(bd_i)-j]}  \Big] \\
           &=&\frac{1}{4}\Big[n+  \sum_{i=1}^n\sum_{y=1}^3i^{y[Tr(bd_i)]}\cdot i^{-jy} \Big ] \\
           &=&\frac{1}{4}\Big[n+\sum_{y=1}^3 \chi(ybD)i^{-jy} \Big ],
        \end{eqnarray*}
where $\chi$ is the canonical additive character of ${\cal R}$,
$ybD$ denotes the set $\{ybd:d\in D\}$, and $\chi(ybD)=\sum\limits_{x\in D}\chi(ybx)$
for any subset $S$ of ${\cal R}$.
When $j=0$, then $ N_b(0)=\frac{1}{4}\big[n+\sum\limits_{y=1}^3 \chi(ybD)\big]$.
When $j=2$, then $N_b(2)=\frac{1}{4}\big[n+\sum\limits_{y=1}^3 \chi(ybD)(-1)^{y}\big ]$.
Hence,
\begin{eqnarray*}
  w_L(c_b) &=& n-N_b(0)+N_b(2) \\
   &=&n- \frac{1}{4}\Big[n+\sum_{y=1}^3 \chi(ybD)\Big]+\frac{1}{4}\Big[n+\sum_{y=1}^3 \chi(ybD)(-1)^{y}\Big ] \\
  &=& n-\frac{1}{4}\Big[\sum_{y=1}^3 \chi(ybD)-\sum_{y=1}^3 \chi(ybD)(-1)^{y}\Big ]\\
  &=&n-\frac{1}{2}[\chi(bD)+\chi(-bD)],
\end{eqnarray*}
that is,
\begin{eqnarray}\label{EQ1}
  w_L(c_b) &=&n-\Re(\chi(bD)).
\end{eqnarray}

\section{Construction of $\Z_4$-codes by Boolean functions}

Let $f$ be a Boolean function from $\bbF_{2^m}$ to $\bbF_{2}$. The
support of $f$ is defined to be
$$S_f=\{x\in \bbF_{2^m}:f(x)=1\}\subseteq \bbF_{2^m}.$$
Take a set $\bar{S}_f\subseteq {\cal T} $ such that $\bar{S}_f\equiv S_f~(\mathrm{mod}~2)$.
Let the size of set $S_f$ be $n_f$, that is, $n_f=|S_f|=|\bar{S}_f|.$ The Walsh-Hadamard transform of $f$ is defined by
\begin{equation}\label{EQ2}
 W_{f}(w)=\sum_{x\in \bbF_{2^m}}(-1)^{f(x)+tr(wx)},
\end{equation}
where $w\in \bbF_{2^m}$. By \cite{D}, the Walsh spectrum of $f$ is the following
multiset $$\{\{ W_{f}(w):w\in \bbF_{2^m}    \}\}. $$

\subsection{Case of $D=\{d: d\in \bar{S}_f\}$}

Defined $\Gamma(w)=\sum\limits_{x\in {\cal T}}i^{Tr(wx)}$ with $w\in {\cal R}$. In this case $D=\bar{S}_f$, the weight distribution of
$C_{\bar{S}_f}$ can be worked out in this subsetion. To this end, we need the following lemma \cite{YHKS}.

{\lem Let $\epsilon$ be the primitive 8th root of unity, given by $\epsilon=\frac{(1+i)}{\sqrt{2}}$. For any $w=r+2s\in {\cal R}$ with $r(\neq0),s\in{\cal T}$, we have
$$\Gamma(w)=i^{-Tr(\frac{s}{r})}\Gamma(1) $$
with $ \Gamma(1)=\begin{cases}
\sqrt{2^m}\epsilon^m,~~~~~if~m~is~odd,\\
-\sqrt{2^m}\epsilon^m,~~~if~m~is~even.
\end{cases}$
}

Now, define $Q(x)=\sum\limits_{\substack {i,j=0\\j>i}}^{m-1}x^{2^i+2^j}$ and $\hat{f}(w)=2^{-m}\sum\limits_{x\in {\cal R}}i^{2f(\bar{x})+Tr(wx)}$, where $w\in {\cal R}$ and $f(\bar{x})=1$ if $x\in \bar{S}_f$, otherwise 0.
The function $\hat{f}(w)$ is {\bf not} the Walsh-Hadamard transform but it plays a similar role in estimating the weight distribution as the classical Walsh-Hadamard transform in \cite{D}. Moreover, the main result of this subsection is described in the following
theorem.

{\thm Let symbols and notations be as above. Let $w=r+2s\in {\cal R}$ with $r,s\in{\cal T}.$ Then $C_{\bar{S}_f}$ is a linear code over $\Z_4$
with length $n_f$ and its Lee weight distribution
is given by the following multiset:
\begin{eqnarray}
\left\{\left\{ \frac{ 4n_f-2\Re(\Gamma(w))+\Re(W_{f_r}(\bar{s})+W_{f_r}(\bar{r}+\bar{s}))}{4} \right \}\right\}&\cup&
 \left\{\left\{ \frac{2n_f+\Re(W_f(\bar{s}))}{2} \right \}\right\}\\
 &\cup& \{\{ 0\}\}\nonumber ,
\end{eqnarray}
with $f_r(x)=f(x)+Q(rx )$.}
\begin{proof} It is trivial that $w_L(c_0)=0$.
As defined previously, we have $2^m\hat{f}(w)=\sum\limits_{x\in {\cal R}}i^{2f(\bar{x})+Tr(wx)}$, where $ f(\bar{x})=1$ if $x\in \bar{S}_f$, and $f(\bar{x})=0$ if $x\notin\bar{S}_f$. Let $w=2s\in I$ with $s\in{\cal T}\backslash \{0\}.$ From the previous discussion in the Sections 2.1 and 3, we have

$$ 2^m\hat{f}(w)=\sum_{\substack {x=y+2z\in {\cal R}\\ x\in {\cal T}+2{\cal T}}}i^{2f(\bar{x})+Tr(wx)}
=2^m \sum_{ y\in {\cal T}}(-1)^{f(\bar{y})+tr(\bar{s}\bar{y})}=2^mW_f(\bar{s}).$$
On the other hand,
\begin{eqnarray*}
  2^m\hat{f}(w) &=& \sum_{\substack {x=y+2z\in {\cal R}\\ x\in {\cal T}+2{\cal T}}}i^{2f(\bar{x})+Tr(wx)} \\
  &=& \sum_{\substack {x=y+2z\\ x\in\bar{S}_f+2{\cal T}}}i^{2f(\bar{x})+Tr(wx)}+
  \sum_{\substack {x=y+2z\\x\in ({\cal T}\backslash\bar{S}_f)+2T}}i^{2f(\bar{x})+Tr(wx)}\\
  &=&\sum_{z\in {\cal T}}\sum_{y\in\bar{S}_f}i^{2f(\bar{y})+Tr(wy)} +
   \sum_{z\in {\cal T}}\sum_{y\in T\backslash\bar{S}_f}i^{2f(\bar{y})+Tr(wy)}\\
   &=&2^m
   \Big( - \sum_{y\in\bar{S}_f}i^{Tr(wy)} +\sum_{y\in {\cal T}\backslash\bar{S}_f}i^{Tr(wy)}  \Big)\\
  &=& -2^{m+1}\chi (w\bar{S}_f),
\end{eqnarray*}
where the last equality follows by $\sum\limits_{y\in {\cal T}}i^{Tr(wy)}=\sum\limits_{y\in \F_{2^m}}(-1)^{tr(\bar{s}\bar{y})} =0$ with $s\neq0$.
It then follows from $(1)$ that the Lee weight of the codeword
$c_w$ with $w\in I\backslash \{0\}$ is equal to $ \frac{2n_f+\Re(W_f(\bar{s}))}{2}$.

It remains to consider $w=r+2s$ with $r\neq0$. Now consider $E(w)=\sum\limits_{d\in \bar{S}_f}i^{Tr(wd)}=\sum\limits_{x\in {\cal T}}\frac{1-(-1)^{f(\bar{x})}}{2}i^{Tr(wx)}$, which implies $w_L(c_w)=n_f-\Re(E(w))$. So it is necessary to analyze the exponential sum $E(w)$. By a simple calculation, we have
\begin{eqnarray*}
  2E(w) &=& \sum_{x\in {\cal T}}i^{Tr(wx)}-  \sum_{x\in {\cal T}}(-1)^{f(\bar{x})}i^{Tr(wx)} \\
   &=& \Gamma(w)- \sum_{x\in {\cal T}}(-1)^{f(\bar{x})+tr(\bar{s}\bar{x})}i^{Tr(rx)} \\
   &=&   \Gamma(w)- \sum_{x\in {\cal T}}(-1)^{f(\bar{x})+tr(\bar{s}\bar{x})+Q(\bar{r}\bar{x})}i^{tr(\bar{r}\bar{x})}
\end{eqnarray*}
where the last equality follows by $Tr(rx)=tr(\bar{r}\bar{x})+2Q(\bar{r}\bar{x} )$ in \cite{KH}.
 Furthermore, let $f_r(\bar{x})=f(\bar{x})+Q(\bar{r}\bar{x} )$, we get $\Re(E(w))=\frac{1}{2}\Re(\Gamma(w))-\frac{1}{4}\Re(W_{f_r}(\bar{s})+W_{f_r}(\bar{r}+\bar{s}))$.
Hence, the Lee weight
distribution of $C_{\bar{S}_f}$ is given by the multiset in $(3)$. This completes the
proof. 
\end{proof}

Note that $Q(x)=\sum\limits_{\substack {i,j=0\\j>i}}^{m-1}x^{2^i+2^j}$ is a quadratic function. The rank of the quadratic form $g$ is defined to be the codimension of the binary vector space
$$V_g=\{x\in \F_{2^m}: g(x+z)-g(x)-g(z)=0~\mathrm{for} ~\mathrm{all}~z\in \F_{2^m}\}.$$
Let $g$ be a quadratic form of rank $h$, then $|V_g |=2^{m-h}$.
When $m\leq 8$, it not difficult to determine numerically the rank of $Q(x).$ Indeed it can be proved that the bilinear form $B$ attached to $Q$ is
$$B(x,y)= \sum\limits_{\substack {i,j=0\\j>i}}^{m-1}x^{2^i}y^{2^j}+y^{2^i}x^{2^j},$$
a bivariate polynomial that can be factored in Magma.
Based on this data, we give the following proposition.

{\prop Let $h=\lfloor \frac{m}{2}\rfloor$. Then the rank of $Q(x)$ is equal to $2h$.}

Let $f$ be a affine function. Without loss of generality, let $f(x)=tr(ax)+b$ with $a\in \F_{2^m}^*$ and $b\in \F_2$.
Using Equation (\ref{EQ2}), for any $s\in \F_{2^m}^*$, we have
\begin{equation}\label{EQ4}
  W_f(s) =\sum_{x\in \F_{2^m}}(-1)^{tr(ax)+b+tr(sx)}
   = (-1)^b2^m\delta_{a,s},
\end{equation}
where $\delta_{a,s}$ is equal to 1 if $a=s$, and 0 if not. We are now ready for a connection with affine functions. To this end, we need the following classical lemma \cite[vol. 2, p. 1802]{PH}.

{\lem Let $f$ be a Boolean degree 2 and let $2h$ be the rank of the associated quadratic Boolean function. Let
$$W_f(\lambda)=\sum_{x\in \F_{2^m}}(-1)^{f(x)+tr(\lambda x)},~~~\lambda\in \F_{2^m},$$ then $\{W_f(\lambda):\lambda\in \F_{2^m}\}$ has the distribution $\{0,\pm 2^{m-h}\}$.}

The following corollary introduces a connection between affine functions and a class of
linear code $C_{\bar{S}_f}$ over $\Z_4$. And the Lee weight distribution of the linear
code $C_{\bar{S}_f}$ is established.

{\coro Let $m\geq2$ be a positive integer. Let the rank of $Q(x)$ is equal to $2h$. If $f(x)=tr(ax)+b$ is a affine function in $m$ variables with $a\in \F_{2^m}^*, b\in \F_2$, then $C_{\bar{S}_f}$ is a linear code
with parameters $(n_f,4^m)$ with $n_f=2^{m-1},$ and its Lee weight distribution is $\{0\}\cup\{ 2^{m-1}, 2^{m-1}+(-1)^b2^{m-1}\}\cup\{ 2^{m-1}-\frac{1}{2} \Re(\Gamma(w)),2^{m-1}-\frac{1}{2} \Re(\Gamma(w))\pm 2^{m-h-2}, 2^{m-1}-\frac{1}{2} \Re(\Gamma(w))\pm 2^{m-h-1}\}.$ }

\begin{proof} It is not hard to get $n_f=2^{m-1}$ when $f$ is a affine function.
The result follows by Theorem 4.1, Lemma 4.2 and Equation (\ref{EQ4}).\end{proof}

{\ex For $m=4$ and $b=1$, we have $\Re(\Gamma(w))=\pm4,$ and we obtain the Lee distance $d_L=4.$ After Gray map we obtain a $(16,2^8,4)_2$ code, one unit away from the optimal linear $[16,8,5]_2.$}

{\ex For $m=5$ and $b=1$, we have $\Re(\Gamma(w))=\pm4,$ and we obtain the Lee distance $ d_L=10.$ After Gray map we obtain a $(32,2^{10},10)_2$ code, two units away from the optimal linear $[32,10,12]_2.$}

In the sequence, we will employ bent Boolean functions and semi-bent Bollean function to construct linear binary codes with only a few weights. From Theorem 4.2, we can obtain the following results immediately.

{\coro Let $m>2$ be an even integer and $f$ be a bent Boolean function in $m$ variables, then the torsion code $Tor(C_{\bar{S}_f})$ of the linear code $C_{\bar{S}_f}$ is a linear code with parameters $[n_f,m, \frac{1}{2}(n_f-2^{\frac{m-2}{2}})]_2$, where $n_f=2^{m-1}\pm2^{\frac{m-2}{2}}$. And its weights are $\frac{1}{2}(n_f\pm2^{\frac{m-2}{2}})$.
}

{\ex For $m=4$ we have $n_f=6,$ and we obtain the Hamming distance $ d=2$ of $Tor(C_{\bar{S}_f})$. The best know binary $[6,4]_2$ linear code has distance 2 only. }

{\ex For $m=4$ we have $n_f=8,$ and we obtain the Hamming distance $ d=3$ of $Tor(C_{\bar{S}_f})$. The best know binary $[8,4,3]_2$ linear code is extended Hamming code. }

{\coro Let $m>3$ be an odd integer and $f$ be a semi-bent Boolean function in $m$ variables, then the torsion code $Tor(C_{\bar{S}_f})$ of the linear code $C_{\bar{S}_f}$ is a linear code with parameters $[n_f,m, \frac{1}{2}(n_f-2^{\frac{m-1}{2}})]_2$, where $n_f=2^{m-1}\pm2^{\frac{m-1}{2}}$. And its weights are $\frac{1}{2}(n_f\pm2^{\frac{m-1}{2}})$, and $\frac{1}{2}n_f$.
}

{\ex For $m=5$ we have $n_f=12,$ and we obtain the Hamming distance $ d=4$ of $Tor(C_{\bar{S}_f})$. The best know binary $[12,5]_2$ linear code has distance 4. }


\subsection{Case of $D=\{d=x+2y: x\in \bar{S}_f, y\in {\cal T}\}$}

In the light of Theorem 4.2, the linear code $C_{\bar{S}_f}$ has more Lee-weights. In order to get few Lee-weights, we change the definition set of $C_D$.
In this subsection, the definition set of $C_D$ is $D=\{d=x+2y: x\in \bar{S}_f, y\in {\cal T}\}$. The main result of this subsection is described in the following theorem.

{\thm Let symbols and notations be as above. Let $w=r+2s\in {\cal R}$ with $r,s\in{\cal T}.$ Then $C_{D}$ is a linear code over $\Z_4$
with length $2^mn_f$ and its weight distribution
is given by the following multiset:
\begin{eqnarray}\label{EQ3}
\left\{\left\{  2^{m}n_f \right \}\right\}&\cup&
 \left\{\left\{  2^{m}n_f+2^{m-1}\Re(W_f(\bar{s})) \right \}\right\}\cup \{\{ 0\}\}.
\end{eqnarray}
 }
\begin{proof} It is trivial for $w_L(c_0)=0$. Let $w=r+2s\in {\cal R}$ with $r,s\in{\cal T}$ and $(r,s)\neq(0,0).$ By a simple calculation, we have
\begin{eqnarray*}
  \chi(wD) &=& \sum_{d=x+2y\in D}i^{Tr(wd)} \\
   &=& \sum_{x\in \bar{S}_f}\sum_{y\in {\cal T}}(-1)^{tr(\bar{r}\bar{y})}i^{Tr(wx)} \\
   &=& 2^m\delta_{r,0} \sum_{x\in \bar{S}_f}i^{Tr(wx)},
\end{eqnarray*}
where $\delta_{r,0}=\begin{cases}
1, ~~~if~r=0,\\
0,~~~if~r\neq 0.
\end{cases}$
As defined previously, if $w\in {\cal R}^*$, we get $w_L(c_w)=|D|=2^mn_f$. Now, suppose that $w=2s$ with $s\in {\cal T}\backslash \{0\}$, then
\begin{eqnarray*}
 \chi(wD) &=& 2^m\sum_{x\in \bar{S}_f}(-1)^{tr(\bar{s}\bar{x})}\\
   &=& 2^m\sum_{x\in {\cal T}}\frac{1-(-1)^{f(\bar{x})}}{2}(-1)^{tr(\bar{s}\bar{x})} \\
   &=& -2^{m-1}W_f(\bar{s}).
\end{eqnarray*}
Applying Equation (\ref{EQ1}), for any $w=2s\in I\backslash \{0\}$, the Lee weight of the codeword
$c_w$ is equal to $ 2^{m}n_f+2^{m-1}\Re(W_f(\bar{s}))$.
Hence, the weight
distribution of $C_{D}$ is given by the multiset in $(\ref{EQ3})$. This completes the
proof.
\end{proof}

 Theorem 4.13 established a connection between Boolean functions and a class of linear code over $\Z_4$. In order to determine the weight
 distribution of the linear code $C_D$, it is equivalent to analyze the Walsh spectrum. We are now ready for a connection with bent functions.

 {\coro Let $m>2$ be an even integer. If $f$ is a bent Boolean function in $m$ variables, then $C_{D}$ is a linear code
with parameters $(2^mn_f,4^m),$ with $n_f=2^{m-1}\pm 2^{\frac{m-2}{2}},$ and non-zero Lee weights $2^mn_f\pm 2^{\frac{3m-2}{2}}$ and $2^mn_f.$}
\begin{proof}
The length is estimated as in the binary case \cite[(17)]{D}. Since, by the definition of a bent function, the Walsh transform takes values in $\{\pm 2^{\frac{m}{2}}\},$
the result follows by Theorem 4.13.\end{proof}
Similarly, if $f$ is a semi-bent function, we have, by \cite[Cor. 11]{D}) the analogous result.

{\coro Let $m>3$ be an odd integer. If $f$ is a semi-bent Boolean function in $m$ variables, then $C_{D}$ is a linear code
with parameters $(2^mn_f,4^m,d_L),$ with $n_f=2^{m-1}\pm 2^{\frac{m-1}{2}}.$ And its non-zero Lee weights are $2^mn_f\pm 2^{\frac{3m-1}{2}}$ and $2^mn_f,$ respectively.}\\



\section{Construction of $\Z_4$-codes by skew sets}
A subset $D$ of $\F_q^*$ is called a skew set of $\F_q$ if $D$, $-D$ and $\{0\}$ form a partition of $\F_q$. Recall that the maximal ideal of Galois ring ${\cal R}$ generated by 2. For simplify, denoted by $I$ the maximal ideal. Now, we extend the notion of skew sets from fields to rings
as follows.

{\Def A subset $D$ of ${\cal R}^*$ is called a skew set of ${\cal R}$ if $D$, $-D$ and $I$ form a partition of ${\cal R}$.}\\

In the light of Definition 5.1, we see that $| D|=| -D|=(2^m-1)2^{m-1}$, which implies the length of $C_D$ is $(2^m-1)2^{m-1}$. We are now ready to prove the following theorem.

{\thm  Let $D$ be any skew set of ${\cal R}$. Then the code $C_D$ is a two-Lee weight code of length $(2^m-1)2^{m-1}$ and its Lee weight distribution is given in Table I.}
\begin{center}$\mathrm{Table~ I. }~~~\mathrm{Lee~weight~ distribution~ of}~C_D $\\
\begin{tabular}{cccc||cc}
\hline
  Weight&&   & & Frequency  \\
  \hline
 0        & &   & & 1\\
    $2^{2m-1}$        & &   &              &$2^{4m}-1$\\
  $(2^m-1)2^{m-1}$  &    & &       &$2^m-1$ \\
  \hline
\end{tabular}
\end{center}

\begin{proof}
In the case $b=0$, the codeword $c_b$ is zero-codeword, thus $w_L(c_b)=0$. It remains to investigate the Lee weight of $c_b$ with $b\neq0$. By the orthogonality relation of character, for any $b\neq0,$ we have
$$\sum_{x\in {\cal R}}\chi(bx)=\chi(bD)+\chi(-bD)+\chi(bI)=0.$$
Note that $$\chi(bI)=\sum_{x\in I}i^{Tr(bx)}=\sum_{x\in \F_{2^m}}(-1)^{tr(\bar{b}x)}=\begin{cases}
0,~~~~\mathrm{if}~\bar{b}\neq0,\\
2^m,~~\mathrm{if}~\bar{b}=0,
\end{cases}$$
 with $b\equiv \bar{b}~(\mathrm{mod}~2)$. Applying Equation (\ref{EQ1}), we get $w_L(c_b)=2^{2m-1}$ or $(2^m-1)2^{m-1}$ depending on the value of $\bar{b}$.
\end{proof}

In the light of Theorem 5.2, $\phi(C_D)$ is a $(2^{2m}-2^{m},2^{2m},2^{2m-1}-2^{m-1})$ binary code. Generally speaking, by using Gray map, the binary image of a linear code over $\Z_4$ is nonlinear. Below we mention a open problems.

{\open Determine the linearity of $\phi(C_D).$}

A central problem of coding theory is to determine the minimum value of $n$, for which
an $(n, M, d)_q$-code or an $[n, k, d]_q$-linear code exists. We denote by $N_q(M, d)$ the minimum
length of a nonlinear code over $\F_q$, with $M $ codewords and distance $d$,
while we
use $L_q(k, d)$ in the case of a linear code of dimension $k$ with distance $d$. Observe that
\begin{equation}\label{eq2}
  N_q(q^k,d)\leq L_q(k,d)
\end{equation}
Recall the Griesmer bound (\cite{G}) on the parameters of an $[n,k,d]_q$ code.

{\lem All $[n,k,d]_q$ linear codes satisfy the following bound
$$n\geq  L_q(k,d)\geq \sum_{i=0}^{k-1}\Big\lceil \frac{d}{q^i} \Big\rceil.$$}

If there exist a $[2^{2m}-2^{m},2m,2^{2m-1}-2^{m-1}]_2$ linear code, we know that $ L_2(2m,2^{2m-1}-2^{m-1})=2^{2m}-2^{m}$ by using Griesmer bound. Consider the case $m=2$, then we have a $[12,4,6]_2$ linear code.

In fact, there is a version of the Griesmer bound for binary non linear codes \cite[Theorem 6]{BGMS} that applies in case of distance of the form $2^a-2^b.$ Thus, combing Lemma 5.4 and Equation (\ref{eq2}), $\phi(C_D)$ is a optimal code.

Besides the binary image $\phi(C_D)$, there exits a binary code which is canonically associated with $C_D.$ In the sequel, we will investigate the binary code which call torsion code $Tor(C_D).$
 {\coro Keep the conditions in Theorem 5.2. Then $Tor(C_D)$ is a $[2^{2m}-2^{m},m,2^{2m-2}]_2$ linear code which meets Griesmer bound.
 }
 \begin{proof}
 It is not difficult to obtain the paraments of $Tor(C_D)$. Applying Lemma 5.4, we obtain the results immediately.
  \end{proof}

{\rem Recall that the non-zero codewords of the $[2^{m}-1,m,2^{m-1}]_2$ simplex code all have weights $2^{m-1}.$ In fact, $Tor(C_D)$ is equivalent to the $2^{m-1}$ copies of $[2^{m}-1,m,2^{m-1}]_2$ simplex code.
}
%
%
%
%

%
%
%
%
%
%
%
%
%

\section{Conclusion}
In the present work we have studied a family of linear codes over the ring $\Z_4.$ Given a classical
Boolean function, or, equivalently its support we have discussed two different constructions of Trace codes. Building on the results in \cite{D}, we give upper and lower bounds
on the minimum Lee distance. It is worth noticing that both the residue and torsion codes of the first type of codes we have constructed here are the trace codes of \cite{D}. This gives a pair of crude bounds on the Lee minimum distance. However, the weight distribution is difficult to analyze in general, and we could give only partial results when the Boolean function is affine. In particular, computing the rank of a certain quadratic form is a challenging open problem.
A better understanding of this quadratic form could lead to a better choice of the Boolean function.
The second construction is easier to analyze, and yields three-weight codes. Eventually, a natural generalization of skew sets yields a family of optimal two-weight codes. Using difference sets or relative difference sets might give more powerful results.
Another path of inquiry would be to use a similar approach for $\Z_4$ valued Boolean function like in \cite{ST}.

{\bf Acknowledgement:} This manuscript is supported by NSFC of China (61202068), Technology Foundation for Selected Overseas Chinese Scholar, Ministry of Personnel of China (05015133) and the Open Research Fund of National Mobile Communications Research Laboratory£¬Southeast University (2015D11).
The authors thank Claude Carlet for helpful discussions on Boolean functions.

\end{document}